\documentclass{SVProc}

\usepackage[utf8]{inputenc}
\usepackage{color}
\usepackage{amsmath}
\usepackage{mathtools}
\usepackage{amsfonts}
\usepackage{subcaption}
\usepackage{epsfig}
\usepackage{epstopdf}
\usepackage{makeidx} 

\makeindex

\usepackage{url}

\newtheorem{Example}{Example}

\begin{document}

\title{On the use of energy tanks for robotic systems}
\date{January 2022}
\titlerunning{On the use of energy tanks for robotic systems}  
%
\author{Federico Califano\inst{1} \and Ramy Rashad\inst{1} \and Cristian Secchi\inst{2} \and Stefano Stramigioli\inst{1}}
\authorrunning{Federico Califano et al.} 
\institute{Robotics and Mechatronics department, University of Twente, The Netherlands\\
	\email{\{f.califano;r.a.m.rashadhashem@utwente;s.stramigioli\}@utwente.nl}
	\and
	Department of Sciences and Methods for Engineering, \\University of Modena and Reggio Emilia, Italy\\
	\email{cristian.secchi@unimore.it}}

\maketitle

\vspace{-7mm}
\begin{abstract}
In this document we describe and discuss \textit{energy tanks}, a control algorithm which has gained popularity inside the robotics and control community over the last years. This article has the threefold scope of i) introducing to the reader the topic in a simple yet precise way, starting with a throughout description of the energy-aware framework, where energy tanks find their genesis; ii) summarising the range of applications of energy tanks, including an original reflection about different formulations of those; iii) discussing limits and future challenges involving energy tanks and energy-aware control in general.   
\end{abstract}

\section{Introduction and Background}
In order to understand in an insightful way the idea behind energy tanks, we have to introduce its originating system theoretic fields, i.e., \textit{port-based modelling} and \textit{passivity-based control}. We first present a motivational introduction collecting the relevant conceptual features and then introduce the basic mathematical notation needed to fully understand the ideas and the techniques behind energy tanks.  

The port-based modeling framework \cite{FILIPPO1991565} relies on \textit{energy}, one of the fundamental concepts in engineering practice and physics. Consistently with a thermodynamic approach, dynamical systems are understood and modelled as devices able to transform and route physical energy across the network they create. They do so by means of \textit{power ports}, representing the entities interconnecting separate systems, and by means of which physical systems exchange energy.
Since energy is the physical quantity which is shared among all physical domains, this paradigm yields a great success in modeling multi-physical systems and has one of its most fortunate outcomes in the \textit{port-Hamiltonian} (pH) system theory \cite{Duindam2009ModelingSystems}. 
In such theory, the classic Hamiltonian modeling framework is extended exploiting bond-graph concepts and port-based thinking in order to lead to a system representation where energy flows are evident and power ports are exploited for representing the power flowing through the system.

Furthermore, pH system are closed with respect to power preserving interconnections. This means that when multiple pH systems are interconnected by their power ports and when no energy is either produced nor dissipated in the interconnection (i.e. energy is just exchanged), the interconnected system is still pH and its properties can be easily inferred from its generating parts, e.g. the total energy of the interconnected system is given by the sum of its constituent ones.

It becomes thus possible to model increasingly large networks in which energy transfers between subsystem are systematically monitored and possibly used for control purposes. 
This framework differs from most control theoretic approaches in which dynamical systems are seen as signal-processing devices. In fact, in the so-called \textit{control by interconnection} method, peculiar of the pH framework, the controller itself is treated as a virtual physical system instead of a signal processor, with its own power ports and pH structure. In this way the closed-loop system, generated by the power preserving interconnection of the pH plant and the pH controller, benefits of maintaining a pH structure as well, that can be used to develop novel control techniques. 

These techniques make an important contact with the branch of nonlinear control theory of \textit{passivity-based control} \cite{vanderSchaftL2}. Passivity is a system theoretic property which, roughly speaking, guarantees that a dynamical system does not generate an unbounded amount of energy autonomously. This property, which is stronger than stability, is a desirable one for the closed-loop system for different and sometimes controversial reasons, which we will discuss in detail later, and which constitute the real essence of the motivation behind energy tanks.

\subsection{Passivity and Port-Hamiltonian Systems}

We now introduce the mathematical preliminaries needed to understand the main properties of pH systems and energy tanks. We keep the discussion as simple as possible to convey the main ideas needed for introducing energy tanks, and refer to \cite{vanderSchaftL2,Duindam2009ModelingSystems} for a throughout description of these preliminaries.

A nonlinear affine control system with state $x \in \mathbb{R}^n$ and input-output pair $(u,y) \in \mathbb{R}^m \times \mathbb{R}^m$ in the form
\begin{align}
\label{eq:nonlin1}
    \dot{x}&=f(x)+g(x)u  \\
    \label{eq:nonlin2}
        y&=h(x)
\end{align}
is said to be passive with respect to a non negative state-dependent storage function $V(x)$ if the following inequality is true for any possible input signal $u(t)$ and at any positive time $t$:

\begin{equation}
\label{eq:passivity}
    V(x(t))-V(x(0)) \leq \int_0^t y^T(s)u(s) ds.
\end{equation}
Passivity implies stability, which is a property of the autonomous system (i.e., $u=0$), under the weak conditions that qualify the storage function as a Lyapunov function \cite{vanderSchaftL2}. Passivity lends itself to a physical analogy: the storage function can be interpreted as a (generalized) energy function and the inner product of the input and of the output as a (generalized) power flow. For these reason, input and output can be called power conjugated variables.

\begin{figure}
	\centering
	\begin{subfigure}{0.65\textwidth}
		\includegraphics[width=\columnwidth]{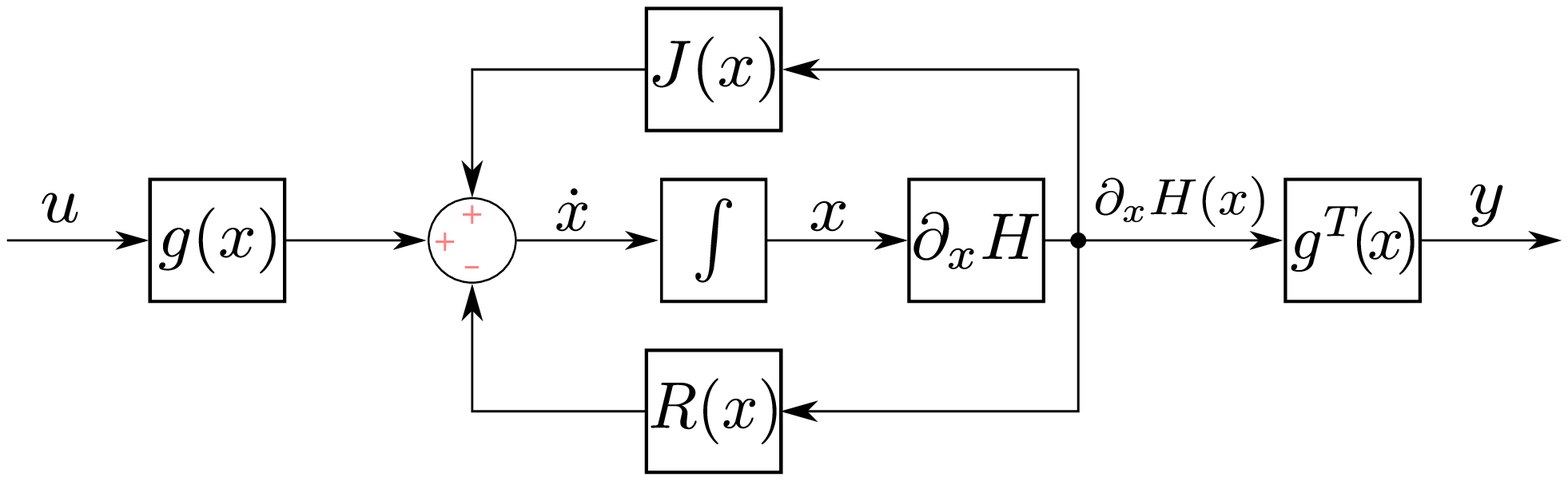}
		\caption{signal-based block diagram}
		\label{fig:pH_sys_BD_signal}
	\end{subfigure}

	\begin{subfigure}{0.35\textwidth}
		\includegraphics[width=\columnwidth]{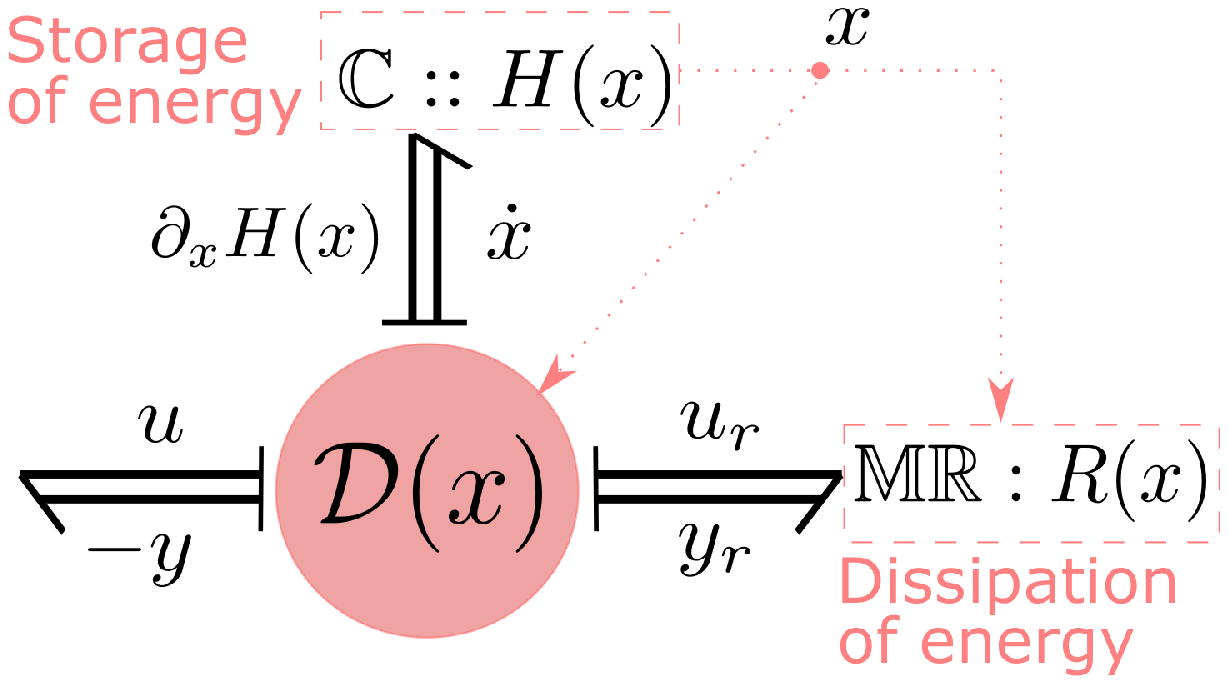}
		\caption{energy-based bond graph}
		\label{fig:pH_sys_BG}
	\end{subfigure}
	$\qquad$
	\begin{subfigure}{0.35\textwidth}
		\includegraphics[width=\columnwidth]{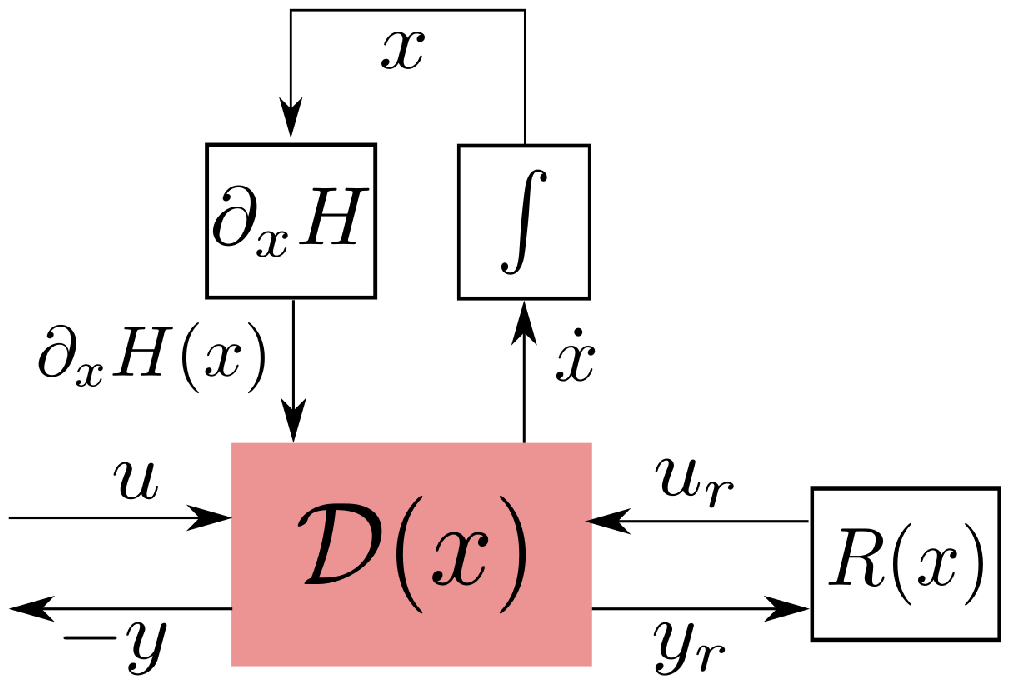}
		\caption{energy-based block diagram}
		\label{fig:pH_sys_BD}
	\end{subfigure}
	\caption{Graphical representation of a generic port-Hamiltonian system}
\end{figure}

The input-state-output formulation of a port-Hamiltonian system is given by the following instance of (\ref{eq:nonlin1}-\ref{eq:nonlin2}):
\begin{align}
\label{eq:pH1}
    \dot{x}&=[J(x)-R(x)]\partial_x H(x)+g(x)u,  \\
    \label{eq:pH2}
        y&=g^T(x) \partial_x H(x).
\end{align}
This structure exhibits explicitly some important geometrical properties of the system, like the \textit{Hamiltonian} $H(x)$, mapping the state variables to the total energy of the system; the symmetric positive semi-definite matrix $R(x)$, representing the energy dissipating elements; and the skew-symmetric matrix $J(x)$, representing the internal interconnection structure along which energy is distributed
and generalising the Symplectic/Poisson structures of analytic mechanics. We indicate with $\partial_x H$ the differential
of the Hamiltonian, represented as a column vector. A graphical representation of the system is depicted in the block diagram of Fig. \ref{fig:pH_sys_BD_signal}.

An immediate property of pH systems (\ref{eq:pH1}-\ref{eq:pH2}) is that they are passive with respect to the Hamiltonian as storage function:
\begin{align}
\label{eq:powerbalancepH}
    \dot{H}&=(\partial_x H(x))^T\dot{x}=-(\partial_x H(x))^TR(x)\partial_x H(x) +y^Tu \leq y^Tu,
\end{align}
where we used the skew-symmetry of $J(x)$ and symmetry and positive semi-definiteness of $R(x)$.
Integrating the latter equation in time yields the passivity condition (\ref{eq:passivity}), which we rewrite to enhance its clear energetic interpretation, indicating with $E_d(t):=\int_0^t (\partial_x H(x))^TR(x)\partial_x H(x) \geq 0$ the dissipated energy:
\begin{equation}
\label{eq:passivitypH}
    H(x(t))-H(x(0)) = \int_0^t y^T(s)u(s) ds - E_d(t) \leq\int_0^t y^T(s)u(s) ds.
\end{equation}
The left hand side of the equation represents the stored energy in the system, which is always less or equal than the supplied energy through the I/O port, represented by the right term. The positive dissipated energy $E_d(t)$ determines the convergence rate to the equilibrium in case of an
uncontrolled system, and can be interpreted as a \textit{passivity margin}, i.e., the bigger the dissipated energy is, the bigger is the margin for which the passivity inequality results satisfied. 

The port-Hamiltonian system of equations (\ref{eq:pH1}-\ref{eq:pH2}) can be restructured to highlight more its constituting energetic subsystems. The system components are now classified into energy storage, energy dissipating and energy supply elements, as depicted in the bond graph of Fig. \ref{fig:pH_sys_BG} and its block-diagram equivalent Fig. \ref{fig:pH_sys_BD}. Central to this formulation of the pH system is the so called \textit{Dirac structure} $\mathcal{D}(x)$, which represents the relation between all the power conjugated variables encoded in the pH system. This relation can be represented by the following skew-symmetric map
\begin{equation}
\label{eq:diracpH}
    \begin{pmatrix}
    \dot{x} \\ y_r \\ -y 
    \end{pmatrix}
    =
    \begin{pmatrix}
    J(x) & -I & g \\ I & 0 & 0 \\ -g^T & 0 & 0
    \end{pmatrix}
    \begin{pmatrix}
    \partial_x H(x) \\ u_r \\ u
    \end{pmatrix},
\end{equation}
where $I$ is the $n$-dimensional identity matrix. This representation unravels the power conjugated variables corresponding to the dissipative port $(u_r,y_r)$, related by means of the algebraic equation $u_r=R(x)y_r$. The latter, together with the second equation in (\ref{eq:diracpH}), can be combined to form (\ref{eq:pH1}). The sum of all the power terms represented by the pairing of the conjugated variables gives zero because of the skew-symmetry of the map, i.e.,
 $(\partial_x H(x))^T \dot{x} + y_r^T u_r - y^T u=0$,
which encodes exactly the power balance (\ref{eq:powerbalancepH}) once the previously described algebraic relation on the dissipative port is used.

\subsection{Passivity-Based and Energy-Aware Control}

When it comes to control, port-Hamiltonian theory makes close contact with passivity-based control, a set of techniques aiming at designing a controller in such a way that the closed-loop system is passive, directly achieving stability \cite{vanderSchaftL2}. This objective is automatically encoded in the pH structure, since the form (\ref{eq:pH1}-\ref{eq:pH2}) is conserved under power-preserving interconnection of pH systems. As a consequence, the passive system (\ref{eq:pH1}-\ref{eq:pH2}) can represent both the open-loop system (i.e., the uncontrolled plant) and a \textit{target} closed-loop system, with desired and physically interpretable properties. The control objective becomes then to design a pH controller such that the desired target pH closed-loop system is achieved. Instances of these techniques are the \textit{energy balancing} (EB-PBC) method, in which the target Hamiltonian function is assigned; and the \textit{interconnection damping assignment} method (IDA-PBC), in which all the system matrices of the target system are assigned \cite{Duindam2009ModelingSystems}. 
In what follows we elaborate on the motivation for achieving a closed-loop passive system in this framework.

\begin{figure}[t]
	\centering
	\begin{subfigure}{0.5\textwidth}
		\includegraphics[width=\columnwidth]{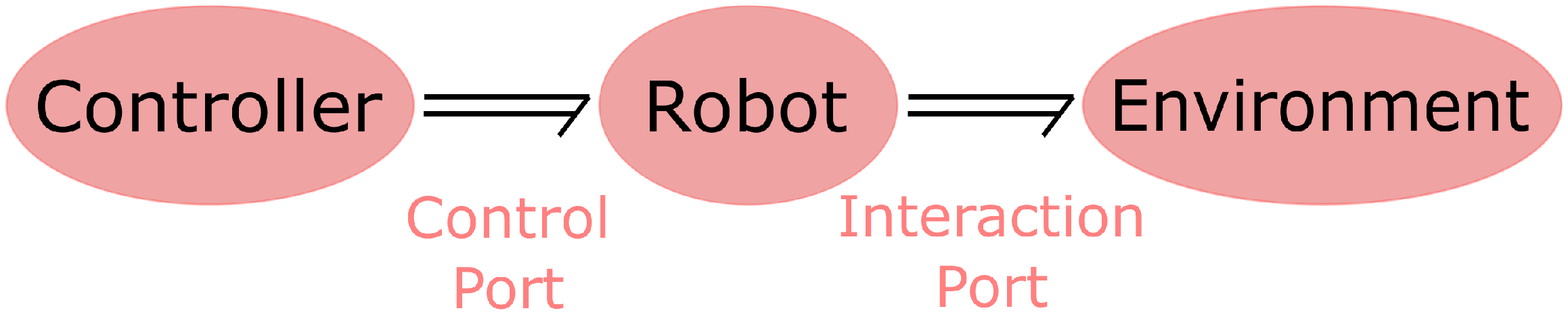}
		\caption{passivity-based control}
		\label{fig:pbc_architect}
	\end{subfigure}
	\begin{subfigure}{0.6\textwidth}
		\includegraphics[width=\columnwidth]{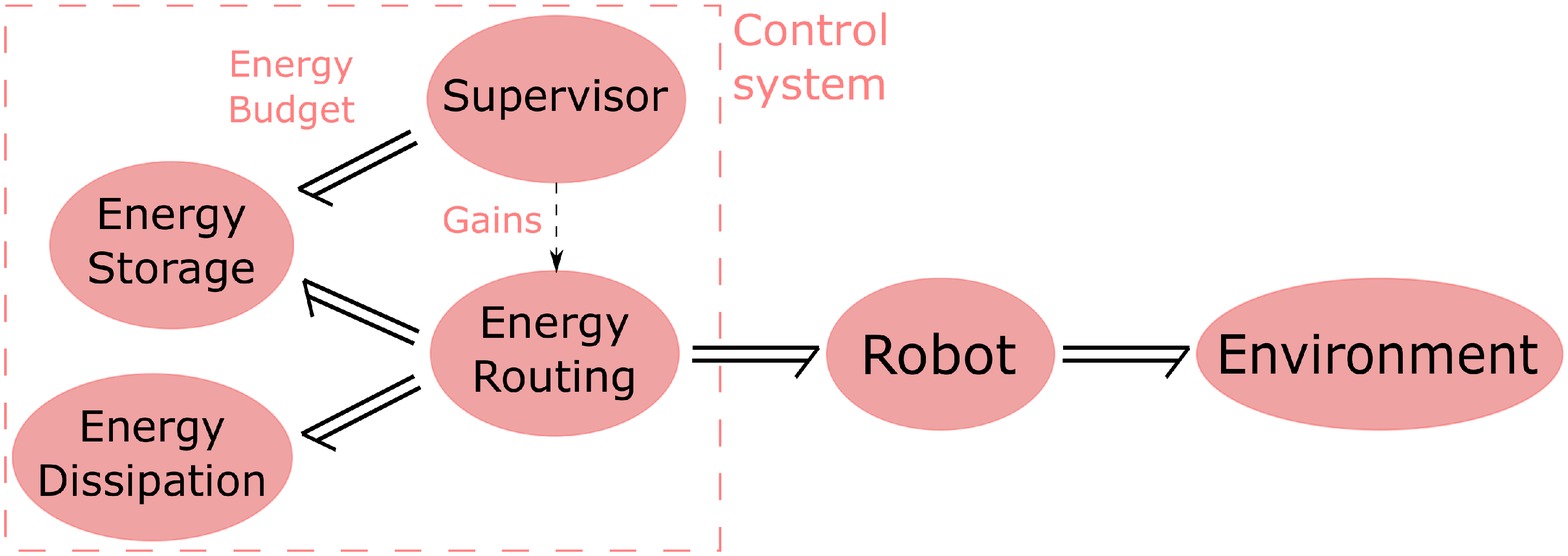}
		\caption{energy-aware control}
		\label{fig:eac_architect}
	\end{subfigure}
	\caption{Comparison between the standard passivity-based control and energy-aware control frameworks}
\end{figure}

A physical system, like e.g., a robot manipulator, which will be taken as archetypical example in the following, can be understood as a pH system with two distinct ports: one connecting the mechanical system to an unknown environment (which is another physical system), called interaction port; and one connecting it to its actuators, commanded by the control system to be designed (represented as a pH system as well), called control port, as shown in Fig. \ref{fig:pbc_architect}.
The described passivity-based control techniques in the pH setting have the objective of shaping the systems' dynamics by means of the controller such that the closed-loop system exhibits a new, desired behavior at the interaction port. The achievement of stability in this context is just a particular case of possible closed-loop dynamics, i.e., the case in which the interaction port is disconnected. 
Along the task it executes, the robot will interact with the environment, whose dynamics is often unknown.
Contrarily to what is often done in e.g., robust control theory, making specific simplifying assumptions on the structure of the environment as a dynamical system (like linearity) is considered misleading and conceptually wrong. What matters for safety and performance purposes is a stable interconnection between the system and its environment along the task execution. The interaction will follow physical laws, dependent on the unknown parameters of the environment. The following claim, referred to as \textit{passivity as must}, is taken as a primary objective in this framework:
\textit{The controlled robot should result in a passive behaviour seen from the environment port.} The motivation behind this claim lies on the fact that if the controlled robot is passive, then the interconnection with any passive environment would indeed result in a stable interconnection. Furthermore, as formally proven in \cite{Stramigioli2015Energy-AwareRobotics}, it is always possible to construct a passive environment such that its interconnection with a non passive controlled system results in an unstable behavior. 
This set of considerations made passivity of the controlled system a necessary condition in the field.

Designing control systems with passivity as the sole goal is often considered over-conservative and non optimal in terms of performance. In fact, it is no surprise that a design which is completely agnostic with respect to an environment that the system has to interact with, is questionable in terms of performance of a task, which potentially comprises an interaction with the environment itself. Furthermore some tasks (e.g., periodic locomotion) require a continuous injection of energy in the system to be executed, and cannot be achieved with a pure passive design, which we stress serves at the purpose of stability of interconnection with a passive environment.
On the other hand, the \textit{energy-aware control framework} \cite{Stramigioli2015Energy-AwareRobotics} extends the standard passivity-based framework with a set of modules such that the goal is not primarily achieving passivity, but being able to take decisions observing the energy flows happening insight the whole control network.
With reference to Fig. \ref{fig:eac_architect}, a fundamental constituent of the energy-aware framework is \textit{energy routing}, first introduced in \cite{Duindam2004Port-basedSystems}. This technique introduces the possibility to design nontrivial power-preserving interconnections between two or more systems such that the energy could be transferred from one system to another without violating passivity \cite{Stramigioli2015Energy-AwareRobotics}. These tools, which can be used together as abstractly represented in Fig. \ref{fig:eac_architect} to form an energy aware control system, serve in linking a low-level passivity based control framework to one more aware of performance and integrated at a decision level. 

The energy routing technique is fundamental for directly controlling the energy flow and, consequently for implementing a desired dynamic behavior for the controlled system. Starting from the operating principles outlined in \cite{Duindam2004Port-basedSystems}, the \textit{energy tank} has been firstly proposed in \cite{Secchi2006PositionTelemanipulation}. Using energy tank it is possible to passively implement any desired control action according to the available energy budget \cite{Stramigioli2015Energy-AwareRobotics}.
The energy budget is provided by a high-level control module referred to as the \textit{supervisor} in Fig. \ref{fig:eac_architect}. This supervisor module
is interconnected to the control system and can provide to it new energy if needed. In this perspective, if the supervisor does not provide new energy to the controlled system, then the passivity requirement allows flows of energy coming only from the environment. The supervisor implements a task-dependent \textit{energy budget protocol}, which serves to optimally perform the task.

\section{Energy tanks}
\begin{figure}
	\centering
	\begin{subfigure}{0.72\textwidth}
		\includegraphics[width=\columnwidth]{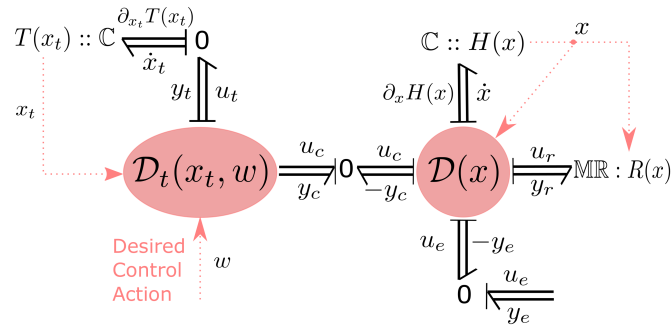}
		\caption{bond graph}
		\label{fig:etank_basic_BG}
	\end{subfigure}
	\begin{subfigure}{0.72\textwidth}
		\includegraphics[width=\columnwidth]{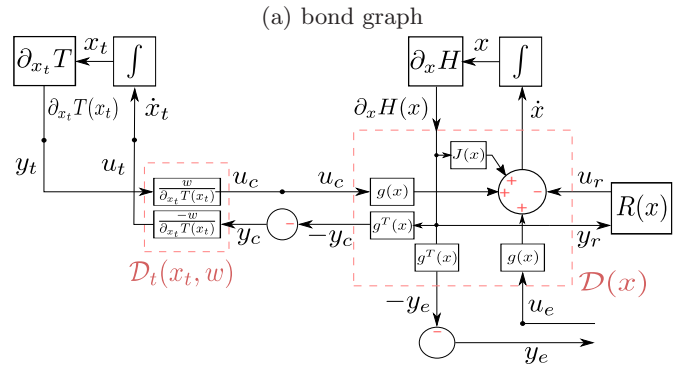}
		\caption{block diagram}
		\label{fig:etank_basic_BD}
	\end{subfigure}
	\caption{Basic energy tank control implementation graphically represented by bond graphs (a) and block diagrams (b).}
\end{figure}

Mathematically an energy tank is a dynamical system which constitutes an atomic energy storing element. Using a pH formulation (\ref{eq:pH1}-\ref{eq:pH2}), and indicating the variables with a subscript $t$, it can be represented as
\begin{align}
\label{eq:tank1}
    \dot{x}_t&=u_t  \\
    \label{eq:tank2}
        y_t&= \partial_{x_t} T(x_t),
\end{align}
where $T(x_t)$ is the non negative energy function for the tank. This is just a simple integrator with possibly nonlinear output, and is used to store information about the current energy budget in the system. In particular the goal is to interconnect the tank (\ref{eq:tank1}-\ref{eq:tank2}) to the plant (\ref{eq:pH1}-\ref{eq:pH2}), in a way that allows the implementation of some control action which fulfills the execution of some task, and at the same time meeting the desired passivity constraint. This is possible thanks to a suitable power-preserving interconnection between the two systems, exploiting the previously discussed idea of energy routing, i.e., the lossless energy transfer between the tank and the plant. As described in the last section, let us consider the pH plant (\ref{eq:pH1}-\ref{eq:pH2}) distinguishing between the interaction port (with input $u_e$) and the control port (with input $u_c$). Mathematically this is achieved by setting $u=u_c+u_e$, and as a consequence both the control and the interaction input are conjugated to the same original output $y_c=y_e=y$. 
The power preserving interconnection between the plant and the tank reads formally
\begin{equation}
\label{eq:interconnection}
    \begin{pmatrix} u_c \\ u_t \end{pmatrix} = \begin{bmatrix} 0 & \frac{w}{\partial_{x_t} T(x_t)} \\ -\frac{w^T}{\partial_{x_t} T(x_t)} & 0\end{bmatrix} \begin{pmatrix} y_c \\ y_t \end{pmatrix},
\end{equation}
where $w$ is the desired task-dependent control action to be passively implemented. We represent the interconnection of the systems using both block diagrams and bond graphs, respectively in Fig. \ref{fig:etank_basic_BD} and \ref{fig:etank_basic_BG}, where we abstractly represent the interconnection as a state-modulated structure $\mathcal{D}_t$. Notice that $\mathcal{D}_t$, due to its skew-symmetry, is a power preserving interconnection. In fact, it can be easily seen that $u_c^Ty_c=-u_ty_t$. Nevertheless, $\mathcal{D}_t$ is not a Dirac structure since it is modulated by the co-energy variable $\partial_{x_t}T(x_t)$. This has profound implications on the geometric structure of the controlled system, that is no more port-Hamiltonian but that can be modeled as a port contact system, as discussed in \cite{SECCHI2007474}.

This interconnection produces two key effects, whose combination embodies the role of energy tanks. First, it correctly implements the desired action, i.e., from the side of the plant (\ref{eq:pH1}), one obtains
\begin{equation}
    \dot{x}=[J(x)-R(x)]\partial_x H(x)+g(x)w+g(x)u_e.
\end{equation}
Secondly, the interconnection is power-preserving, which can be inferred from the skew-symmetry of the matrix in (\ref{eq:interconnection}). To show this explicitly let us consider the closed-loop system, with closed-loop energy $\mathcal{H}:=H+T$:

\begin{align}
\label{eq:closedlooppH}
    \begin{pmatrix} \dot{x} \\ \dot{x}_t \end{pmatrix} &= \begin{pmatrix} J(x)-R(x) & g(x) \frac{w}{\partial_{x_t} T(x_t)} \\ -\frac{w^T}{\partial_{x_t} T(x_t)} g^T(x) & 0\end{pmatrix} \begin{pmatrix} \partial_x  \mathcal{H} \\ \partial_{x_t} \mathcal{H} \end{pmatrix}+\begin{pmatrix} g(x) \\ 0 \end{pmatrix} u_e\\
    y_e &= \begin{pmatrix}
    g^T(x) & 0
    \end{pmatrix} \begin{pmatrix} \partial_x  \mathcal{H} \\ \partial_{x_t} \mathcal{H} \end{pmatrix}
\end{align}

Evaluating the variation of the closed-loop Hamiltonian along the system trajectories one obtains
$\dot{\mathcal{H}}= (\partial_x \mathcal{H})^TR(x)\partial_x \mathcal{H}+y_e^Tu_e \leq y_e^Tu_e$,
which proofs passivity of the closed-loop system with respect to the interaction port $(u_e,y_e)$, with $\mathcal{H}$ as storage function, and with the same dissipation rate of the original system. This means that the (undefined in sign) power $y^T w$ flowing on the port of the original system is at any time exchanged with the tank without being dissipated or generated. In other words, and considering for simplicity a case in which no natural dissipation ($R(x)=0$) and no external interaction ($u_e=0$) is present for the original system, one obtains $\dot{\mathcal{H}}=0$, or equivalently $\dot{T}=-\dot{H}$,
which means that the power flowing in/out the tank is equal to the power flowing out/in the original system, i.e., the energy routes from one system to another without being created or destroyed.  

These properties introduce an important advantage in the context of energy-aware control, since the input $w$ is free to be chosen arbitrarily, without the need of being generated by a pH controller representing a passive physical system (e.g., a spring and a damper, resembling the traditional PD controller). The energy in the tank $T(x_t)$ plays the role of the amount energy that is still at disposal of the control mechanism implementing the action $w$ before loosing the described formal passivity. In fact, it is important to notice that the interconnection (\ref{eq:interconnection}), as well as the closed-loop system (\ref{eq:closedlooppH}) become singular when $\partial_{x_t}T(x_t)=0$, representing the moment at which it becomes impossible to passively perform the desired action $w$. This passivity property manifests with respect to the closed-loop storage function $\mathcal{H}$, which in turns depends on the choice of the energy tank function $T(x_t)$. Let us consider the most common choice of the latter, namely $T(x_t)=\frac{1}{2} x_t^2$, for which $\partial_{x_t} T(x_t)=x_t$. In this case, in order to meaningfully implement the energy tank algorithm, it is needed to continuously observe the energy in the tank and its variation in order to implement a control action $\alpha w$, rather than $w$, where

\begin{equation}
\label{eq:alphavalve}
     \alpha= 
\begin{cases}
    1,& \text{if } (T\geq \epsilon \geq 0) \,\, \textrm{or} \,\, (T< \epsilon \,\, \textrm{and} \,\, \dot{T}>0 )\\
    0,& \text{if } (T< \epsilon \,\, \textrm{and} \,\, \dot{T} \leq 0 ),
\end{cases}
\end{equation}

for some small positive energy level $\epsilon$. In this way formal passivity of the closed-loop system is recovered completely since the system is just detached from the tank and the controller in the moment in which no energy budget is left, possibly having to wait till the supervisor replenishes the tank.
A less understood fact is that the interconnection (\ref{eq:interconnection}) might fail to generate a well-posed closed-loop system even for choices of $T(x_t)$ which analitically exclude the condition $\partial_{x_t}T(x_t)= 0$. It is in fact tempting, as proposed in \cite{Zheng2018Time-varyingTank}, to choose as energy tank function $T(x_t)=e^{x_t}$, and thus avoid the aforementioned singularity condition and the construction in (\ref{eq:alphavalve}). Nevertheless this choice produces new problems, like the fact that the closed-loop storage function $\mathcal{H}$ would not qualify as Lyapunov function for the origin of the closed-loop system, renouncing to formal stability. An instance of an even subtler problem with this choice is illustrated in the next example, and is to the best of our knowledge not discussed in the literature.

\begin{Example}
\label{ex:example}
Consider a simple mass subject to a force $F$, with kinetic energy $H(p)=\frac{1}{2m}p^2$, momentum $p:=mv$, being $v$ the velocity of the mass. The pH equations of motion are
$\dot{p}=F=u_c;
    y_c=\partial_p H(p)=v.$
Let us consider the interconnection (\ref{eq:interconnection}) with the tank (\ref{eq:tank1}-\ref{eq:tank2}) using $T(x_t)=e^{x_t}$. The initial conditions are set as $p(0)=0$ (i.e., $H(0)=0$), and $x_t=0$ (i.e., $T(0)=1$). Let us perform a task in which we constantly accelerate the mass, applying a constant force $\Bar{F}$, i.e., we set $w=\Bar{F}$ in (\ref{eq:interconnection}). As usual $-\dot{H}=\dot{T}$ holds, and thus the time evolution of the energy in the tank satisfies $T(t)=T(0)-\int_0^t \dot{H} \textrm{dt}=1-\Bar{F}\int_0^t v \textrm{dt}=1-\frac{\Bar{F}^2 t^2}{2m}$ where we used the explicit solution for the velocity $v(t)=\frac{\Bar{F} t}{m}$. We notice that since $T\geq0$ by definition, the interconnection (\ref{eq:interconnection}) fails to work at $\Bar{t}=\frac{\sqrt{2m}}{\Bar{F}}$. To see this let us consider the tank dynamics (\ref{eq:tank1}), where the input is $u_t$ is determined by the second equation in (\ref{eq:interconnection}). The equation reads $\dot{x}_t=-\frac{\Bar{F}v}{e^{x_t}}=-\frac{\Bar{F}^2  t}{m e^{x_t}}$, where we substituted again the time dependent solution for the velocity. The latter differential equation in $x_t$ possesses a solution which escapes to $-\infty$ at the previously mentioned finite time, making the whole tank algorithm and any closed-loop analysis invalid for a wider time interval.
\end{Example}
\begin{figure}[]
	\centering
	\includegraphics[width = \textwidth]{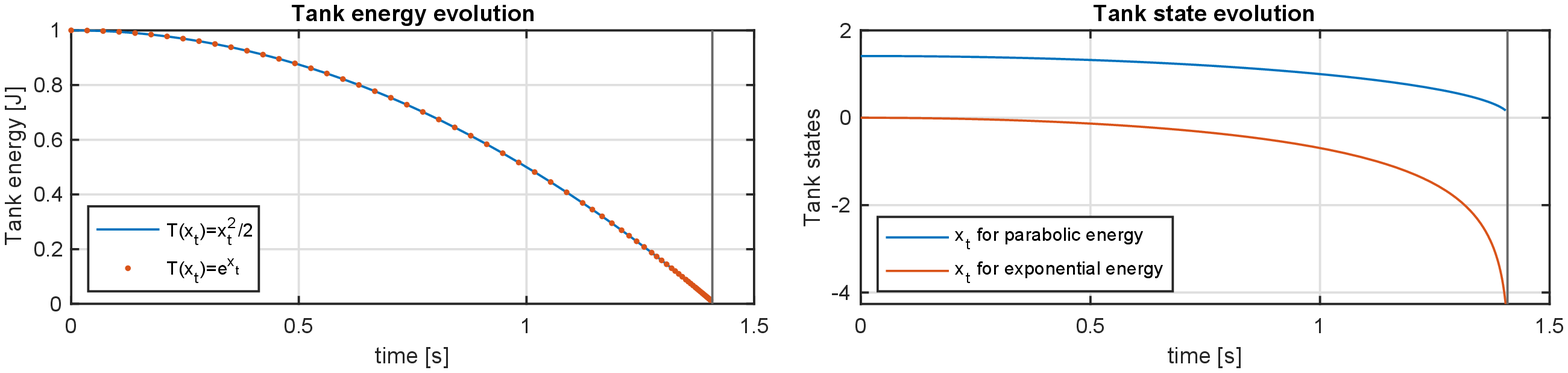}
	\caption{Simulations of tank energy and tank states (with initial conditions chosen to have the same initial energy) for parabolic and exponential choice of $T(x_t)$ for Ex. \ref{ex:example}, with $m=1$, $\Bar{F}=1$.}
	\label{fig:tankcomparison}
\end{figure}
This simple example is very instructive to show that special care must be taken when implementing the interconnection (\ref{eq:interconnection}), and that the exponential choice for the energy of the tank does not prevent the singularity since $\partial_{x_t}T(x_t)= 0$ could happen anyhow due to a finite-time escape of the state in (\ref{eq:tank1}). Furthermore, since $-\dot{H}=\dot{T}$ holds for any form of $T(x_t)$, and since $\dot{H}$ does not depend on the form of $T(x_t)$, the evolution of the tank energy in time does not depend on the specific form of $T(x_t)$.
As a consequence, the specific choice of energy $T(x_t)$ corresponds to a specific choice of coordinate for the energy variable $x_t$, and does not influence any objective energy behaviour. In other words, using $T(x_t)=e^{x_t}$ (resp. $T(x_t)=x_t^2/2$) means just using a coordinate that "observes" a zero energy at $x_t=-\infty$ (resp. at $x_t=0$). To conclude, stating that a specific choice of $T(x_t)$ avoids the singularity in (\ref{eq:interconnection}) is only a coordinate-based illusion: the singularity is only "pushed" at $-\infty$ in the chart where one reads the $x_t$ state in the exponential version. This is shown in Fig. \ref{fig:tankcomparison}, where the simulations corresponding to the task described in Ex. \ref{ex:example} are reported for both choices of the tank energy.
We conclude that, unless one finds numerical advantages in the digital implementation of one specific version of $T(x_t)$, the quadratic choice is one that does not cause a loss of theoretical generality, and the conditional check in (\ref{eq:alphavalve}) must be performed in any case to preserve passivity.

\subsection{Extensions and Applications}

In the aforementioned discussion, the basic energy tank (\ref{eq:tank1}-\ref{eq:tank2}) has been introduced as a simple integrator with one power port $(u_t,y_t)$ connected to the plant via (\ref{eq:interconnection}) with the initial energy budget $T(x_t(0))$ being the only control parameter to be specified by the user.
Many extensions of this basic energy tank implementation have been introduced in the literature and is an active area of research. In what follows we present some of these extensions.

One common extension is to exploit (some of) the dissipated energy of the controlled system (i.e., $E_d$ in (\ref{eq:passivitypH})) to refill the energy tank in a controlled manner  \cite{Dietrich2017PassiveTanks,Rashad2019EnergyApproach,Rashad2022EnergyFramework}.
Another extension is to put an upper bound on the allowable stored energy in the tank which prevents storage of very high virtual energy that could potentially lead to unsafe behavior between the robot and its environment. Such extension, implemented e.g. in \cite{Dietrich2017PassiveTanks,Rashad2022EnergyFramework},  was realized by the addition of a virtual overflow valve that would start dissipating the power flowing to the tank once the energy level inside reaches a user-specific maximum.

The idea of using virtual valves to implement high level control objectives using energy tanks was further extended in \cite{Shahriari2018Valve-basedObjectives} by considering not only the energy level of the tank but also by observing the amount of extracted power and limiting it if needed to prevent possible unsafe interaction scenarios \cite{Tadele2014CombiningRobots}. 
An application of this was demonstrated in \cite{Rashad2022EnergyFramework} where high spikes in the energy tank's power act as signals for possible contact-loss with the environment.

Another common extension is to implement the discontinuous tank detachment law (\ref{eq:alphavalve}) in a smooth manner, e.g. as in \cite{Shahriari2018Valve-basedObjectives}, to avoid introducing chattering in the control signals. Furthermore, one can incorporate high level control-objectives by detaching the tank not only when the energy level depletes but in other scenarios such as when a contact-loss is detected with the environment as in \cite{Schindlbeck2015UnifiedTanks}. In the recent work of \cite{Shahriari2020PowerTanks}, an adaptive extension to energy tanks was proposed to deal with dynamic passive environments relying on varying the power limits of the energy tanks online through an iterative learning procedure.

In conclusion, the above non-exhaustive list of extensions demonstrates that energy-aware control using energy-tanks is an active research topic. Furthermore, it is being utilized in a wide variety of application domains in robotics as summarized in Table .

\begin{table}[]
	\centering
	\begin{tabular}{|l|l|l|l|}
		\hline
		\textbf{Domain} & \textbf{References} & \textbf{Domain} & \textbf{References} \\ \hline
		Aerial physical interaction           & \cite{Rashad2019EnergyApproach,Rashad2022EnergyFramework,Brunner2022EnergyObjectsb}                                  & Robotic surgery                      &        \cite{Ferraguti2013AStiffness,Ferraguti2015AnSurgery,Su2020BilateralConstraint}                                  \\ \hline
		Collaborative robotics                &      \cite{Tadele2014CombiningRobots,Raiola2018DevelopmentRobots}                                    & Shared Control                       &   \cite{Selvaggio2019PassiveGuidance,Selvaggio2016EnhancingGeneration}                                       \\ \hline
		Generic manipulation                  &    \cite{Schindlbeck2015UnifiedTanks,Kronander2016PassiveSystems,Dietrich2017PassiveTanks}                                       &  Teleoperation architectures                                    &     \cite{Franken2011BilateralTransparency,Heck2018DirectDelays,Pacchierotti2015EnhancingFeedback}                                     \\ \hline
		Immitation learning                   &     \cite{Kastritsi2018ProgressiveControl}                                     &  Teleoperation of aerial robots                                    &    \cite{Mersha2014OnRobots,Zhang2021StableFunctions}                                      \\ \hline
		Multi-robot systems                   &      \cite{Riggio2018OnInterconnection,Shahriari2022Passivity-BasedManipulation,Franchi2012BilateralTopology}                                     & Teleoperation of legged robots                                      &    \cite{Risiglione2021Passivity-basedTime-delays}                                      \\ \hline
		Rehabilitation robotics               &   \cite{Najafi2020UsingRehabilitation}                                        &                                      &                                          \\ \hline
	\end{tabular}
\caption{The use of energy-tanks in different robotic application domains.}
\end{table}

\section{Discussion and Future Challenges}
\label{sec:limitsandfuture}

The mechanism of energy tanks has distinguished itself for elegance and simplicity, providing a very convenient way to formally establish passivity of a controlled system. Although the success of the approach, it also raised some critiques and perplexities, which range from the entire energy-aware paradigm to the more specific implementation of energy tanks.
In what follows we collect some of these issues with the hope to act as catalysts for fruitful discussions and reflections. 
\subsubsection{Passivity as must.} The passivity as must requirement is a claim sitting at the very foundation of the energy-aware framework, and it is worth reflecting on its real significance since it is absolutely not a commonly shared necessary condition in the control community. The requirement is in no way connected to a performance metric for the execution of a task. Instead it is based on the general commonsense-based rule that it is wrong to assume any model for the environment to study the stability of the closed-loop system when interfacing the unknown environment along the task execution. Only with a passively controlled system we can indeed infer stability of the interconnection with any passive environment. Although the argument is logically valid, it is worth questioning if, in the context of the execution of a complex task, the distinction between a passive and an active environment is significant in the design process. 

The framework tends to focus on the environment as a system rather that on optimising performance over a task. It is therefore legitimate to question under which conditions it makes sense to start with the passivity as must requirement, reflecting on the relation between classes of tasks and classes of environments.

\subsubsection{Safety} There is often a confusion between the concepts of passivity and safety. Indeed often it is naively said that passivity is necessary for safety, but this is not the case. First of all, when considering the interaction with the environment, safety is not a mathematically defined system theoretic property, and depends on the class of tasks, of systems and on the specific safety hazards that are being considered. The formal passivity condition for the closed-loop system interconnected with energy tanks does not avoid practically unsafe behaviours, especially for high energy in the tank $T(x_t)$. In fact there is no guarantee that the high (yet bounded) amount of energy in the tank is extracted in a short time interval, leading to dangerous power ejections from the system. This aspect is problematic with the use of energy tanks in their basic form, since both high initialisation of energy and a possible refilling of the tank along the task execution can enhance the problem.
The safety aspect has been treated with the use of energy tanks in e.g., \cite{Raiola2018DevelopmentRobots,Tadele2014CombiningRobots}, without removing some perplexity. Indeed, in
these works the initially passive controller is integrated with a safety protocol, which happens to destroy passivity. Motivated by the passivity as must claim, this situation is then recovered by means on energy tanks, used to passivise the control action commanded by the safety protocol. In this case the passivity as must assumption seems to create circular patterns in the design of the control strategy, and the underlying unanswered question remains: why should one chose to use energy tanks to passivise a control action which is already designed to be safe? 
\subsubsection{Need of energy budgeting} Once a complex task has to be executed in an energy-aware control framework, the supervisor has the role of injecting, following a task-dependent protocol, the energy to achieve the task. This is where passivity breaks down, and the more abstract concept of energy-awareness comes into the game. From an energy tank perspective the energy budgeting protocol should respond to the questions: What is the initial energy budget in the tank? Under what circumstances does the tank need to be refilled? In general, in order to assure a good performance along the task, complex energy budgeting protocols need to be implemented in a case by case fashion. This procedure becomes more involved for complex tasks, and may again raise the question under which class of tasks this is a framework worth to stick with from the beginning.

\subsubsection{Towards Energy-Aware Optimisation}

To recap, the energy-aware framework is founded on a claim imposing the necessity of a controlled system which is passive at the environmental port, independently on the task to be performed. This can be always achieved using the basic energy tank architecture. There is no standard way to address the problem of performance optimisation and energy budgeting in this framework, which become particularly critical in situations in which the task cannot be accomplished passively (e.g., periodic locomotion).
The safety aspect appears transversally as a third, separated issue to be solved, additionally to passivity of the internal controller and optimisation of the performance along the task. These issues are deeply intertwined, and need to be addressed together.

A task-depended solution to this passivity-safety-performance problem represents the main open challenge in this energy-aware framework. Energy tanks are perfect candidates to be used is this context since they: i) can passivise any control action; ii) encode information about energy content and power flows undergoing the system, making them extremely useful to deal with safety and energy efficiency metrics; iii) allow for the implementation of arbitrary control actions, for which optimisation over a performance (and/or safety) metric is very tempting.
As possible modification of the mentioned objective, we stress that energy tanks should act as energy and power observers, and not be blindly used to recover passivity of the controlled system, which is ultimately the main reason why perplexities raised in the framework. This could lead to a relaxation of the passivity as must claim, in favour of an energy-aware solution achieving safety and performance optimisation along a task execution.
Preliminary works that addressed the usability of energy tanks in this direction include \cite{Shahriari2018Valve-basedObjectives,Shahriari2020PowerTanks,Capelli2022PassivityEnergy,califano22,benzi22}.

\section{Conclusions}
In this paper we discussed the genesis of energy tanks in the context of its containing system theoretic frameworks, that are port-based modeling and energy-aware control. In addition to listing extensions of the basic energy tank implementation and its range of applications, we raised a series of discussion topics involving passivity, safety, and performance optimisation in relation with the use of energy tanks, with the goal of clarifying the future challenges in establishing this control technique for robotic applications.
 
\bibliographystyle{plain}
\bibliography{referencesFinal.bib}

\end{document}